\documentclass{aa}
\input{epsf.tex}
\begin{document}

   \thesaurus{03(11.05.1; 11.19.2; 11.12.2; 11.03.4 Coma (=Abell 1656)}

\title{$UV$ (2000 \AA) luminosity function of Coma cluster galaxies}

   \author{S. Andreon}


   \institute{Osservatorio Astronomico di Capodimonte, via Moiariello 16,
80131 Napoli, Italy (email: andreon@na.astro.it)}

   \date{Received ... accepted ...}

   \maketitle

   \begin{abstract}

The $UV$ (2000 \AA) luminosity function (hereafter $UV$ LF) of Coma cluster
galaxies, based on more than 120 members, is computed as the 
statistical difference between counts in the Coma direction and in
the field. Our $UV$ LF is an up-date of a preliminary constrain on the 
$UV$ LF previously computed without the essential background counts. The $UV$
LF is well described by a power law with slope $\alpha\sim0.46$, or
equivalently, by a Schechter function with $M^*$ much brighter than the
brightest cluster galaxy and with a slope $\alpha_S\sim-2.0$ or larger. 
In spite of what happens in the optical band, low luminosity
galaxies give a large contribution to the integral luminosity, and by
inference, to the total metal production rate. Galaxies blue in $UV-b$
and/or $b-r$ dominate the Coma cluster $UV$ LF, both in number and 
luminosity. The major source of error in the estimate of the $UV$ LF cames
from the background determination in the Coma direction, which is still
uncertain, even though constrained at high and low amplitudes by  
redshift surveys covering the studied field.


\end{abstract}

\section{Introduction}

In spite of the darkness of the sky at ultraviolet wavelengths (O'Connell
1987) and of the crucial role played by the $UV$ emission in the
determination of the metal production rate, the $UV$ band is still one of
the less explored spectral regions.  This is even more true for objects
in the local Universe, because non redshifted
$UV$ emission can be observed only from space.
In both the single stellar population and continuous star formation scenarios,
the $UV$ luminosity of late--type galaxies
appears to be largely dominated by young massive
stars, thus implying a direct link between $UV$ luminosities and
star formation rates (e.g. Buzzoni 1989). 

In recent years, the understanding that samples of galaxies at very high
redshift can be selected from multicolor deep images (such as the {\it
Hubble Deep Field}), has renewed the interest in $UV$ observations,
allowing tentative determinations of the $UV$ luminosity function
(hereafter LF) for galaxies at $z>2$ (Steidel et al. 1999, Pozzetti et al.
1998). In the local Universe, available samples of $UV$ data for normal
galaxies are generally not suitable for these types of studies due to
either the lack of well defined selection criteria (see for instance the 
IUE sample
reviewed in Longo \& Capaccioli 1992) or to the optical selection of the
objects.  Exceptions to this rule are the samples produced by the FOCA
experiment (Milliard et al. 1991) which
allowed to derive, among various other quantities, 
the {\it local field} $UV$ luminosity function
(Treyer et al. 1998), and to constrain the $UV$ luminosity
function of galaxies in the Coma cluster (Donas, Milliard, \& Laget 1991, 
hereafter DML91). 

In this paper we rediscuss the $UV$ luminosity function (LF, hereafter) of
galaxies in the Coma cluster, first explored by DML91.  
Since DML91 two important sets of data have been acquired: the sample of galaxies
with known redshift in the Coma direction has increased by about 60 \%,
and background counts in $UV$, essential for computing the LF,
have been measured.


The paper is structured as follows: we first describe the data used
(\S 2), then, we present the color--magnitude and color--color relations
for galaxies in the Coma cluster direction (\S 3) and we show that the
availability of colors does not help in identifying
interlopers. In \S 4 we use field counts and the
extensive redshift surveys in the Coma cluster direction to constrain
background counts in the Coma direction and to derive the Coma
cluster LF, presented in \S 5. In \S 6 we discuss the bivariate LF and,
finally, in \S 7 we compare the Coma $UV$ LF to the recently determined field
LF. A summary is given in \S 8. 

In this paper we adopt $H_0=50$ km s$^{-1}$ Mpc$^{-1}$.

\section{The data}

Among nearby clusters of galaxies, Coma ($v\sim7000$ km s$^{-1}$) 
is one of the richest ($R=2$) ones. At a first glance, it looks relaxed and
virialized in both the optical and X-ray passbands.  For this reason it
was designed by Sarazin (1986) and Jones \& Forman (1984) as the
prototype of this class of clusters. The optical structure and photometry
at many wavelengths, velocity field, and X-ray appearance of the cluster
(see the references listed in Andreon 1996) suggest the existence of
substructures. Since these phenomena are also observed in many other
clusters (Salvador-Sol\'e, Sanrom\`a, \& Gonz\'ales-Casado 1993), the Coma
cluster appears typical also in this respect. 

Coma was observed in the $UV$ with a panoramic detector (FOCA). 
Complementary data, are taken from Godwin, Metcalfe \& Peach
(1983; blue and red isophotal magnitudes designed here $b$ and $r$,
respectively) and Andreon (1996; radial velocities taken from
the literature and updated for this paper by means of new NED
entries and accurate morphological types).

The FOCA experiment consisted in a 40-cm Cassegrain telescope equipped
with an ITT proximity focused image intensifier coupled to a IIaO
photographic emulsion. The filter, centered at 2000 \AA \ with a bandwidth
of 150 \AA , has negligible red leakage for objects as red as G0
stars and little dependence of the effective wavelength upon the object
effective temperature. Observations of the Coma cluster were obtained with
a field of view of 2.3 deg and a position accuracy of about 5 arcsec.
The angular resolution of 20 arcsec FWHM was too coarse to allow an
effective discrimination between stars and galaxies (for more details on
the experiment see Milliard et al. 1991). The observations consisted of
many short exposures, totalizing 3000s, and were obtained in April 1988. 
The galaxy catalog and details on the data reduction were published in DML91
and Donas, Milliard, \& Laget (1995, hereafter DML95). 

Coma $UV$ selected sample is found by DML95 to be complete down to $UV\sim17-17.5$
mag and 70\% complete in the range $17.5<UV<18$ mag and includes only $UV$
sources with at least one optical counterpart. Detected objects were
classified by DML91 and DML95 as stars or galaxies according to their
optical appearance. 

Following DML91, the $UV$ magnitude is defined by the expression:  $UV =
-2.5 log(F_{\lambda})-21.175$ where $F_{\lambda}$ is the flux in ergs
cm$^{-2}$A$^{-1}$. Typical photometric errors are 0.3 mag down to
$UV\sim17$ mag and reach 0.5 mag at the detection limit $UV\sim18$ mag.


\section{Color--magnitude and color--color diagrams}

Figure 1 (upper panel) shows the $UV-b$ vs $UV$
color--magnitude diagram for the 254 galaxies detected in the UV in the
Coma field. This sample includes a larger number of galaxies with
known redshift than in DML91, due
to the numerous redshift surveys undertaken since 1991.
We consider as Coma members only galaxies with $4000 <
v < 10000$ km s$^{-1}$ (which is similar or identical to the
criteria adopted by Kent \& Gunn (1982), Mazure et al. (1988),
Caldwell et al. (1993), Carlberg et al. (1994),
Biviano et al. (1995), Andreon (1996), De Propris et al. (1997)).

Figure 1 shows that only a few galaxies detected in $UV$ are near to the
optical catalog limit ($b=21$ mag), except at $UV\sim18$ mag, suggesting
that only a minority of UV galaxies are missed because they are not
visible in the optical\footnote{We stress out that the $UV$ galaxy catalog
contains only sources with at least one optical counterpart, see DML95.}.
This confirms the DML91 statement that the $UV$ sample is truly $UV$
selected, except maybe in the last half--magnitude bin. 

Many optically--faint and $UV$--bright galaxies have not measured redshift. The
lower panel in Figure 1 shows the optical $b-r$ vs $b$ color-magnitude
diagram for the brightest (in $b$) 254 galaxies in the same field. We have
accurate morphological types for all galaxies brighter than $b\sim16.5-17$ mag
(Andreon et al. 1996, 1997). Coma
early--type galaxies (i.e. ellipticals and lenticulars) have $UV-b\sim3$
mag and $b-r\sim1.8$ mag (DML95, Andreon 1996).

The comparison of the two panels in Figure 1 shows several interesting
features. First of all, bright $UV$ galaxies are blue and not
red, as instead is the case in the optical. In other words, early--type
galaxies, due to their $UV$ faintness, do not dominate the $UV$ 
color--magnitude diagram. Red and blue galaxies are small fractions of
the $UV$ and optically selected samples, respectively. 

In second place, galaxies show a much larger spread in  $UV-b$ (7
mag for the whole sample, 6 mag for the redshift confirmed Coma members)
than in $b-r$ or in any other optical or optical--near--infrared color (see,
for example, the compilation in Andreon (1996)). From the theoretical
point of view, such a large scatter in color implies that the $UV$ and $b$
passbands trace the emission of quite different stellar populations. For
all but the very old stellar populations, the $UV$ traces mainly the
emission from young stars (see for instance Donas et al. 
1984; Buat et al. 1989), having maximum main sequence lifetime of a few
$10^8$ years.  Therefore, for star forming galaxies
the $UV$ is a direct measure of the present
epoch star formation rate. Optical data provide instead a weighted
average of the past to present star formation rate. The large scatter in
color therefore implies that galaxies bright in $UV$ are not necessarily 
massive, but more likely the most active in forming stars.


From the observational point of view, this large scatter in color is a
problem, since deep optical observations are needed to derive 
optical magnitudes and hence colors 
(blue galaxies with $UV=18$ mag have $b\sim20-21$ mag) or even for
discriminating stars from galaxies. This
limitation makes difficult to characterize the properties of 
$UV$ selected samples, such as,  for instance, 
the optical morphology (a galaxy with
$UV=18$ mag is bright and large enough to be morphologically classified
only if it is quite red); the redshift (since they are usually measured from
the optical emission or for an optically selected sample); the luminosity
function of galaxies in cluster (the background subtraction is uncertain
because the stellar contribution is difficult to estimate in absence of a
deep optical imaging), etc. Furthermore, it is dangerous to limit the
sample to galaxies with known redshift or morphological type, since,
this would introduce a selection criterion (mainly an optical selection) which has
nothing to do with the $UV$ properties of the galaxies.

Figure 2 shows the color--magnitude diagram for the field in a direction
that in part overlaps the Coma optical catalog provided by Godwin et al. (1983) and 
includes even a few members located in the Coma outskirts. 
Also these data were obtained with FOCA
(Treyer et al. 1998).  Most of the background galaxies have blue apparent
colors, but with a large spread. Almost no background galaxies lay in the
upper-right corner of the graph, i.e. no background galaxy is simultaneously
very red ($UV-b\sim3$) and
faint ($UV\sim17$).  The selection criteria used by Treyer et al.  (1998)
for studying this sample are quite complex and galaxies with missing
redshift (failed or not observed) are not listed, so that it is not
trivial to perform a background subtraction in the color--magnitude plane
(as it is sometimes done in the optical; see, for instance, Dressler et al.  1994). 


The color-color diagram of galaxies in the direction of Coma (Figure 3) has
already been discussed in DML95. But, in our sample, the number of 
galaxies having known membership is larger by 60\%
(from 61 to 99 galaxies). The diagram shows that background galaxies
have colors overlapping those of known Coma galaxies, and, therefore,
it is not of much use in discriminating 
members from non--members. This conclusion is strengthened by
fragmentary knowledge of colors of $UV$--selected samples, which renders
premature to adopt a color selection criterion for
the purpose of measuring the LF.

\section{Evaluation of background counts in the Coma direction}

Since clusters are by definition volume--limited samples, the measure of
the cluster LF consists in counting galaxies in each magnitude bin
{\it after} having removed the interlopers, i.e. galaxies
along the same line of sight but not belonging to the cluster. In general,
interlopers can be removed in three different ways: by determining the
membership of each galaxy throughout an extensive
redshift survey, by a statistical
subtraction of the expected background contamination (see, for instance,
Oemler 1976), or by using color--color 
or color--magnitude diagrams (see for instance, Dressler et al.  1994
and Garilli, Maccagni \& Andreon 1999). 

In our case, the available color informations are not sufficient to
discriminate members from interlopers, and surveys 
in the Coma direction available in literature are not complete down to the
magnitude limit of our sample. Therefore
we were forced to use a hybrid method to estimate and
remove the background contribution. 

Because the
available membership information is qualitatively different for bright and
faint galaxies, we consider them separately. For almost all galaxies
brighter than $M_{UV}=-19.7$ mag, redshifts are available in the literature,
and interlopers can be removed one by one. For fainter galaxies we compute the
LF from a statistical subtraction of the field counts, and, therefore,
the largest source of error may come from possible large background
fluctuations from field to field. 


Milliard et al. (1992) present galaxy counts in three random
fields, measured with the same experiment used to acquire the Coma data. 
One of the pointings is very near in the sky to 
the Coma cluster. The slope is
nearly Euclidean for the total (i.e.  galaxy+stars) counts
($\alpha\sim0.54$) with a small scatter among the counts in the three
directions (roughly 10\%). After removing the stellar contribution,
galaxy counts have again a nearly Euclidean slope, but an amplitude which is
half the previous one.


Dots in Figure 4 show galaxy counts (i.e. $n(m)$)
in the Coma direction (we simply count all galaxies in each bin, open
dots and dashed line in the figure) and the average of the three ``field
directions" (solid dots and solid line).  At magnitudes fainter than $UV=16$ mag, 
galaxy
counts in the Coma direction are {\it lower} than those in directions not
including clusters of galaxies, although errorbars are quite large.
At first sight, this plot is surprising:
clusters are overdensities and therefore counts in their directions should
be higher than field counts. However, this expectation is not necessarily
correct in the $UV$ band. Star formation is inhibited in the high
density environments (Hashimoto et al, 1998, Merluzzi et al, 1999)
and therefore counts in the direction of the cluster can be similar, or even
lower, to counts not having clusters on the line of sight. 
The $UV$ luminosity is, in fact, a poor indicator for the galaxy mass.

Another possible explanation could be related to large errors and large
background fluctuations from
field to field.  We discuss now in depth this point, taking advantage
of the existence of redshift surveys available in the Coma cluster direction.

Figure 5 shows {\it integral} counts (i.e. $n(<m)$). The solid line is the
integral of the solid line in Figure 4, i.e. it gives
is the {\it expected integral} field galaxy counts. All other lines
refer instead to true measurements in the Coma cluster direction. 
The lower solid histogram in Figure 5 is the lower
limit to the background in the Coma cluster direction, computed as the
sum of the galaxies having (known) velocity falling outside of the 
assumed range for Coma. The upper solid histogram is the upper
limit to the background in the Coma cluster direction, given, instead,
as the sum of the galaxies outside of the assumed velocity range
and the galaxies with unknown membership.
The dotted lines are the $1\sigma$ confidence contours, for the lower and
upper limits, computed according to Gehrels (1986). They simply account
for Poissonian fluctuations and show how large (or small) the real
background could be (at the 68\% confidence level) in order to
observe such large (or small) counts.  A background
lower than the lower dotted histogram would produce too few (at the 68\%
confidence level) interlopers in the Coma cluster direction with respect
to the observed ones; whereas a background higher than the higher dotted
histogram would imply (always at the 68 \% confidence level)
a number of galaxies larger than the size of the sample
(once the Coma members are removed). To summarize, in order 
to be consistent (at the 68\% confidence level)
with redshift surveys in the Coma direction, background counts in the Coma
direction must be bracketed in between the two dotted histogram. 
Assuming smooth counts of nearly Euclidean slope, 
we consider the most extreme amplitudes for the background that 
are still compatible at the 68\% confidence level with the two dashed histograms 
in at least one magnitude bin, and in what follow
we call them  ``maximum$+1\sigma$" and ``minimum$-1\sigma$". 
Under the hypothesis of a nearly Euclidean slope,
the background in the Coma direction turns out to be between
2.8 and 17.8 times smaller than the expectation shown by the line in
Figure 5.  The expected field counts (i.e. the
line in Figure 5) are $\sim3\sigma$ away from the maximum background
allowed by the Coma redshift survey 
(i.e. the upper solid histogram). This is an
unlikely but not impossible situation, in particular when we take into
account that the stellar contribution has been assumed and not
measured in two background fields and that counts are slightly 
over--estimated, due to the presence of the Coma cluster and supercluster
(Treyer et al. 1998) in the last field.

\section{$UV$ Luminosity Function}

In the previous section we derived
an estimate for the background in the Coma cluster
direction or, to be more precise, a range for the amplitude of background counts
under the further assumption of nearly Euclidean slope for background
counts. We can, therefore, statistically remove the background contribution
and compute the faint end of the LF (at bright magnitudes
the membership is known for each individual galaxy) and look at the
dependence of the LF on the assumed values of the background amplitude.
Therefore, the determination of the faint end of the LF still depends in
part on the poorly known background counts, but much less than in DML91 
since at that time the slope and the amplitude of
the background contribution were almost unknown and 
it was left free to span 
over a range extending from almost all the data to zero. 

In order to clarify the error implied by our limited knowledge of the 
background counts, we compute twice the lower end of the LF, assuming a
minimum$-1\sigma$ background and a maximum$+1\sigma$ one.  
The actual Coma $UV$ LF is bracketed in between.


We made use of a maximum--likelihood method (Press et al. 1992)
to fit the differential LF of
Coma with a Schechter (1976) or power law functions: 

$ f(m)=\phi^* \ 10^{0.4 (\alpha_S+1)(m^*-m)} \ exp(-10^{0.4(m^*-m)})$

$ f(m) = k \ 10^{\alpha m} $ 

The most important advantage of the maximum--likelihood method is that it
does not require to bin the data in an arbitrarily chosen bin size and
works well also with small samples where the $\chi^2$ fitting is not
useful. It also naturally accounts for lower limits (bins with zero counts
if data are binned). 

The maximum likelihood method leaves the normalization factor undetermined
(since it is reduced in the computation). We therefore derived it by
requiring
that the integral of the best fit is equal to the observed number of
galaxies. In our case we have 125 and 233 galaxies in the Coma sample,
depending on the adopted background subtraction. 


The Coma cluster $UV$ LF - the first ever derived for a cluster - is shown in
Figure 6. Error bars are large, and only the rough shape of the LF can be
sketched. 

The Coma $UV$ LF is well described by a power law (or alternatively by a
Schechter function with a characteristic magnitude $M^*_{UV}$ much
brighter than the brightest galaxy): a Kolmogorov-Smirnov test could not
reject at more than 20\% confidence level the null hypothesis that the
data are extracted from the best fit (whereas we need a 68\% confidence
level to exclude the model at $1\sigma$). The best slope is
$\alpha=0.42\pm0.03$ and $\alpha=0.50\pm0.03$ assuming a maximum$+1\sigma$
and minimum$-1\sigma$ background contamination, respectively.  In terms of
the slope of the Schechter (1976) function $\alpha_S$, these values are
$-2.06$ and $-2.26$ respectively. The exact value of the background
amplitude, once bound by redshift surveys, have small impact on the
slope of the LF, which is quite steep. The Coma $UV$ LF is
steeper than the optical
LF, ($\alpha_S\sim-1.0$, from 5000 \AA \ to 8000
\AA \ Garilli, Maccagni \& Andreon 1999), when computed within a similar
range of magnitudes (i.e. at $M_3+3$, where $M_3$ is the magnitude of the
3th brightest galaxy of the cluster). 


It needs to be stressed, however, that the computed slope of the LF
depends on the assumption of a nearly Euclidean
slope for galaxy counts (the amplitude is constrained by the redshift
survey). We now measure the effect of neglecting this hypothesis.

A very low limit to the slope of the Coma LF can be computed
under the extreme assumption that all galaxies not confirmed as Coma members
(i.e.  all galaxies without known redshift and those with redshifts
outside the velocity range of Coma members) are actual interlopers. The resulting
LF is shown in Figure 7.  No matter how large and how complex the shape
of background counts in the Coma direction is, this estimate provides
the very low limit to the slope of the Coma LF because 
galaxies with unknown membership are only faint, and they
could only rise the faint part of the LF. In such an extreme case, we find 
$M^*_{UV}=-22.6$ mag, brighter than the brightest cluster galaxy. Fitting
a power law, we find instead $\alpha=0.21\pm0.04$. Even in this case,
however, the slope is larger than what is found in the optical (at
$M_3+3$). This LF is computed with
no assumption about the shape of the background counts.

This LF is unlikely to be near to the ``true" Coma $UV$ LF, because the
assumption that all galaxies with unknown redshift are interlopers is
unrealistic and
implies an over--Euclidean slope ($\alpha\sim0.75$) for the background,
which is much steeper than those observed in the three field
pointings by Milliard, Donas \& Laget (1992).
Nevertheless, this very low limit LF gives the very minimum slope for the
Coma $UV$ LF, $\alpha_S=-1.45$. 

The steep Coma $UV$ LF implies that faint and bright galaxies give
similar contributions to the total $UV$ flux, and that the total $UV$ flux
has not yet converged 4 magnitude 
fainter than the brightest galaxy (or, which is the same, at $M_3+3$). 
Therefore, in order to derive the total luminosity and hence the
metal production rate, it is very important to measure the LF down to 
faint magnitude limits. 

\section{Bivariate LF}


Since the redshift information is quite different for blue ($UV-b<1.7$)
and red ($UV-b>1.7$) galaxies, the two $UV$ LFs are computed in different
ways. Redshifts are available for all the red galaxies (which all belong
to the cluster) and the respective $UV$ LF is easy to compute.
Almost all blue galaxies brighter than $M_{UV}\sim-20$ mag have known
redshift, and therefore the determination of this part of the 
blue LF is quite
robust. For the faint part of the blue LF, we adopt an ``average"
background, given as the average normalization between the
maximum$+1\sigma$ and minimum$-1\sigma$ backgrounds previously computed. 

The resulting bivariate color--luminosity function is given in Figure 8. 
The bulk of the $UV$ emission comes from blue ($UV-b<1.7$) galaxies while
all red galaxies have $M_{UV}>-20$ mag. Therefore, since blue galaxies 
dominate the
$UV$ LF both in number and luminosity, the Coma $UV$ LF is dominated by
star forming galaxies and not by massive galaxies. From previous morphological
studies (Andreon 1996) it turns out that Coma red galaxies in
our sample are ellipticals or lenticulars. The fact that early--type galaxies
contribute little to the $UV$ LF may be explained as a consequence of
the fact that these systems have a low recent star
formation histories. Please note that in the optical, the LF is dominated at
the extreme bright end by the early--type (i.e. red) galaxies (Bingelli,
Sandage \& Tammann 1988, Andreon 1998), and not by blue ones as it is in
$UV$. 

\section{Comparison with the $UV$ field LF}

The $UV$ LF of field galaxies has been recently measured by Treyer et al.
(1998) in a region close to Coma, where they found
$\alpha_S=- 1.62^{+0.16}_{-0.21}$, $M^*_{UV}=-21.98\pm0.3$ mag for a
sample of 74 galaxies. As pointed out by Buzzoni (1998), this slope is
quite different from that assumed for the distant field galaxies by Madau
(1997). 

Is there any significant difference between the Coma cluster and the field
$UV$ LFs?  The best Schechter fit to the field data satisfactorily matches
both the very low limit to the Coma LF and the Coma data after the
subtraction of the maximum$+1\sigma$ background contribution (the
probability of a worse fit is 0.1, according to the Kolmogorov-Smirnov
test, whereas we need a probability of 0.05 to call the fit worse
at $2\sigma$), but does not in the case of minimum$-1\sigma$ background
contribution (the probability of a worse fit is 0.00078, according to the
same test, i.e. the two LF differ at $\sim4\sigma$). 
However, using $\alpha_S-1\sigma$ instead of $\alpha_S$ for
the field LF, the fit to the Coma data cannot be rejected with a
probability larger than 0.02, i.e., the $1\sigma$ confidence contour of
the field LF crosses the $\sim2\sigma$ confidence contour of the Coma LF.
Therefore, given the available data, Coma and field $UV$ LFs are different
at $2-3\sigma$ at most. Given the large errors involved, the field and clusters LFs
result therefore compatible with each other. 

\section{Conclusions}

The analysis of $UV$ and optical properties of Coma galaxies is indicative
of the difficulties encountered in
studying $UV$ selected samples: background
galaxy counts are uncertain (as well as their variance); the background
contamination in the $UV$ color--magnitude plane is poorly known.
In spite of these difficulties we found:

1) galaxies in Coma show a large range of $UV$--optical color (6--7 mag),
much larger than what is observed at other redder passbands. 

2) Blue galaxies are the brightest ones and the color--magnitude relation
is not as outstanding as it is at longer wavebands.  Early--type or red
galaxies are a minority in the Coma $UV$ selected sample. In $UV$, 
the brightest galaxies are the most star forming galaxies and not the 
more massive ones. 

3) In spite of the rather large errors, the 
$UV$ LF discussed here is the first LF ever derived for a cluster.
The major source of error in estimating the $UV$ LF comes from the field
to field variance of the background, that it is subtracted statistically.
Present redshift surveys in the studied
field constrain at high and low amplitudes the background contribution in the
Coma direction, as shown in Figure 5. The Coma $UV$ LF is steep and
bracketed between the two estimates shown in Figure 6, with a likely Schechter slope
in the range $-2.0$ to $-2.3$. Even under the extreme hypothesis that all
galaxies with unknown membership are interlopers, the very minimum slope
of the UV-LF is $\alpha_S\lse- 1.45$. 

4) The steep Coma $UV$ LF implies that faint and bright galaxies give
similar contributions to the total $UV$ flux, and that the total $UV$ flux
has not yet converged 4 magnitude 
fainter than the brightest galaxy (or, which is the same, at $M_3+3$). 
Therefore, in order to derive the total luminosity and hence the
metal production rate, it is very important to measure the LF down to
fainter magnitude limits. 

5) The Coma $UV$ LF is dominated in number and luminosity by blue
galaxies, which are often faint in the optical. Therefore the Coma $UV$
LF is dominated by star forming galaxies, not by massive and large
galaxies.

6) The Coma $UV$ LF is compatible with the field LF at $\sim2-3\sigma$.

\acknowledgements This work has been partially done at the Istituto di
Fisica Cosmica ``G.P.S. Occhialini". Its director, Gabriele Villa, is
warmly thanked for the hospitality. This work would have not been possible
without the good $FOCA$ data and I wish to aknowledge all people involved
in that project for their good work. Jean--Charles Cuillandre, Jose Donas,
Catarina Lobo and, in particular, Giuseppe Longo are
also warmly thanked for their attentive lecture of the paper. The
anonymous referee makes useful comments that help to focus the
paper on its major objectives.

\vfill\eject

\begin{figure}
\epsfysize=8cm
\centerline{\epsfbox[60 190 470 600]{MS8517.f1a}}
\epsfysize=8cm
\centerline{\epsfbox[60 190 470 600]{MS8517.f1b}}
\caption[h]{
Color-magnitude diagrams for galaxies in the Coma cluster direction.
Filled circles are redshift confirmed members (galaxies with $4000<v<10000$
km/s), crosses are redshift confirmed background/foreground galaxies, open
circles are galaxies with unknown redshift. Galaxies to the left of the
dotted line have known morphological type.}
\end{figure}

\begin{figure}
\epsfysize=8cm
\centerline{\epsfbox[60 190 470 600]{MS8517.f2}}
\caption[h]{
Apparent color--magnitude diagram for field galaxies in a direction close to 
Coma. Symbols as Figure 1. Only galaxies brighter than the Coma
catalog limit are displayed.}
\end{figure}

\begin{figure}
\epsfysize=8cm
\centerline{\epsfbox[60 190 470 600]{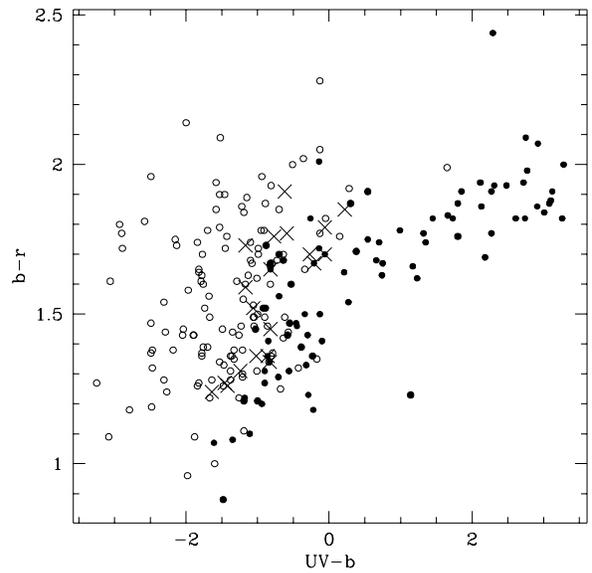}}
\caption[h]{Color--color diagram for galaxies in the Coma cluster direction.
Symbols as in Figure 1.}
\end{figure}

\begin{figure}
\epsfysize=8cm
\centerline{\epsfbox[60 190 470 600]{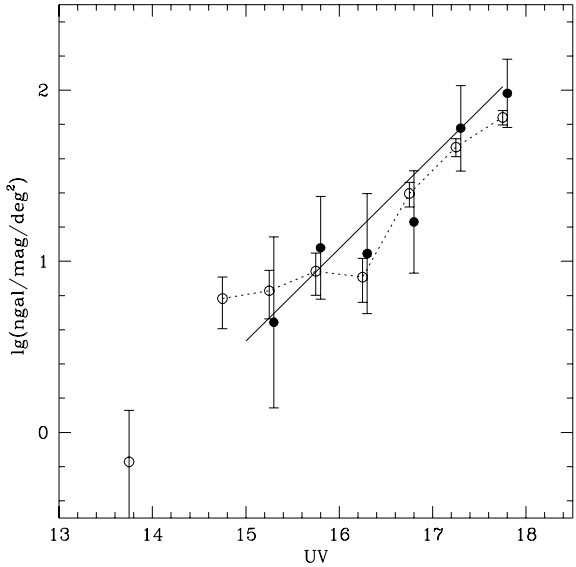}}
\caption[h]{Differential galaxy counts in the Coma cluster direction (open cicles
and dotted line) and in the field (closed points and solid line). 
Error bars of our Coma counts are simply $\pm\sqrt n$.}
\end{figure}

\begin{figure}
\epsfysize=8cm
\centerline{\epsfbox[60 190 470 600]{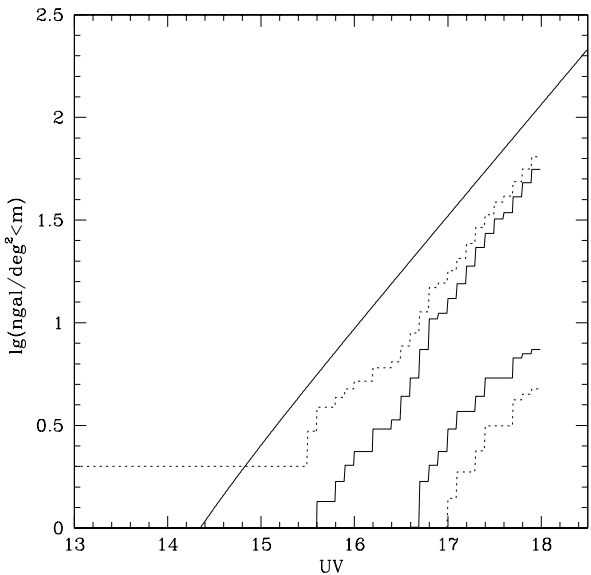}}
\caption[h]{Integrated galaxy counts in different directions:
the expected counts in the field (solid line), the maximum background counts
in the Coma cluster direction (upper solid histogram) and the
minimum background counts (lower solid histogram). Dotted histograms are
$1\sigma$ confidence contours computed according to Gehrels (1986).
At $UV=18$ mag histograms stop because the catalog is limited at
that magnitude}
\end{figure}

\begin{figure}
\epsfysize=8cm
\centerline{\epsfbox[60 190 470 600]{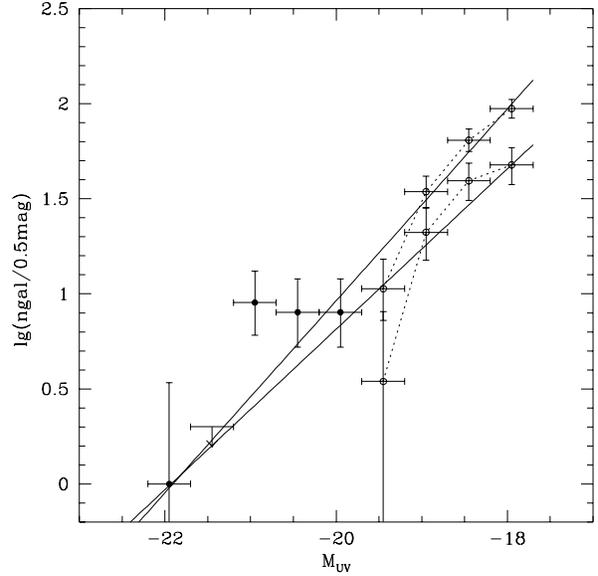}}
\caption[h]{
The $UV$ luminosity function of Coma cluster galaxies. Data in this plot are 
arbitrarily grouped in 0.5 mag bins for presentation purposes only, but in the 
analysis we used non binned data. Error bars in the ordinate direction and 
upper limits are
$\pm1\sigma$ and are computed according to Gehrels (1986).  Error bars in
the abscissa direction show the bin width.  Lines are best fit with
a power law. Details in the text.}
\end{figure}

\begin{figure}
\epsfysize=8cm
\centerline{\epsfbox[60 190 470 600]{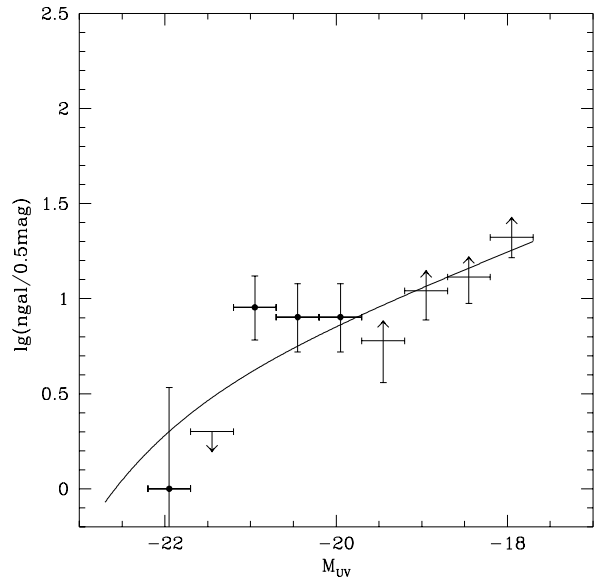}}
\caption[h]{Very low limit to the $UV$ luminosity function of Coma cluster
galaxies. Only redshift confirmed members have been considered and we
have no more assumed a nearly Euclidean slope for background counts.
Data in this graph are arbitrary binned by 0.5 mag for presentation purposes,
but in the analysis we use unbinned data. Errorbars are as in previous figure.
The line is the best fit with a Schechter (1976) function.}
\end{figure}

\begin{figure}
\epsfysize=8cm
\centerline{\epsfbox[60 190 470 600]{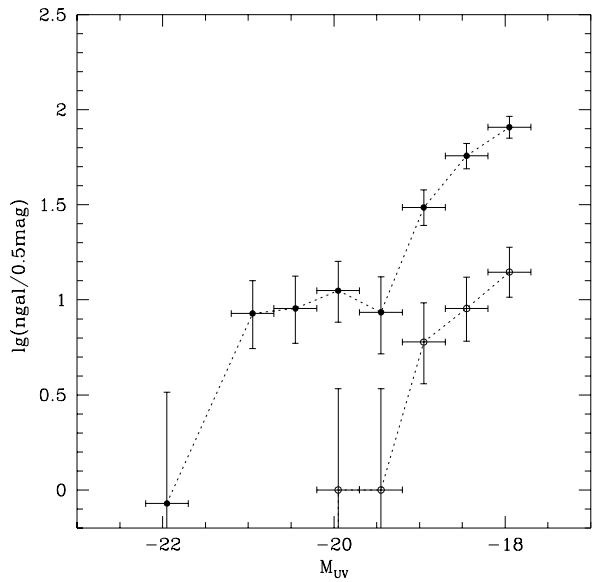}}
\caption[h]{Bivariate luminosity function of Coma galaxies. Close points
refer to blue ($UV-b<1.7$) galaxies, open points refer to red ($UV-b>1.7$)
galaxies. For sake of clarity, upper limits are not drawn. Errorbars are 
as in the previous figure.}
\end{figure}

\end{document}